\documentclass[12pt]{article}
\usepackage[dvips]{graphicx}
\begin{document}
\title{%
Breaking of Nanotube Symmetry by Substrate Polarization
}
\author{%
Alexey G. Petrov$^{1,2}$,
Slava V. Rotkin$^{1,2}$\\
\small \itshape
(1) Beckman Institute, UIUC, 405 N.Mathews, Urbana,
IL 61801, USA;\\
\small \itshape
(2) Ioffe Institute, 26 Politekhnicheskaya st.,
St.Petersburg 194021, Russia.\\
\small \itshape E--mail: rotkin@uiuc.edu
}
\maketitle
\begin{abstract}
Substrate and nanotube polarization are shown to change
qualitatively a nanotube bandstructure. The effect is
studied in a linear approximation in an external potential which
causes the changes. A work function difference between the
nanotube and gold surface is estimated to be large enough to
break the band symmetry and lift a degeneracy of a
lowest but one subband of a metallic nanotube. This subband
splitting for [10,10] nanotube
is about 50 meV in absence of other external
potential.
\end{abstract}

\section{Introduction}
\label{sec:intro}

Since discovery of carbon nanotubes in 1991\cite{iijima},
a deep physics of these
one--dimensional nanoscale objects has been demonstrated.
Fundamental properties of the nanotubes have been studied
in view of possible applications in
electronics and other devices\cite{appl}.
A detailed theoretical description
for electronic structure of ideal single--wall nanotubes
(SWNTs) was obtained as well as for an effect of various
defects and disorder on the SWNT electronic
properties ( \emph{e.g.}, Ref.\cite{louie}). However, the nanotube
systems under experimental study often deviate from a model
picture. In this letter we investigate one aspect of a real
system: the modification of the electronic properties of the SWNT
deposited on a substrate. One expects that symmetry of the
nanotube at the substrate will be lower than the symmetry of the
nanotube itself in vacuum.

A description of a breaking of the symmetry of SWNT
bandstructure due to a charge transfer (or charge injection)
between the nanotube and the substrate (or contacts) and
calculation of a polarization of the substrate and the nanotube,
which follows to the charge transfer, are the goals of
our study. Effects of splitting, mixing, and/or anti--crossing
of the nanotube subbands, that are caused by the depolarization
of the electron charge density, have been almost neglected in
literature before. We use term ``depolarization" for a number of
phenomena including a transverse shift of the electron charge
density from its equilibrium distribution profile (effects
due to an axial/longitudinal depolarization were discussed
elsewhere\cite{jetpl,ecsaluru,tersoff}). We will show that the
\emph{transverse depolarization} results in qualitative changes
of the nanotube density of states (DOS) near van Hove
singularities. In particular, we predict the \emph{splitting} of
a doublet state\cite{damnjan} to be likely observable as a
function of the injected/induced charge density of the SWNT. We
will discuss that in Sec.\ref{sec:model}. In Sec.\ref{sec:fermi}
we will calculate this injected/induced charge density in a
selfconsistent way.

The depolarization and \emph{intrasubband} splitting will be
studied for a typical experimental situation: a single SWNT lies
on a conductive substrate or separated from the conductor by  a
thin insulating layer
%
%
representing an oxide on the surface of a metal. We assume that
the nanotube is connected to electron reservoirs, which may be
the leads or the conductor substrate itself. A transverse
external electric field and/or a work function difference
between the SWNT and the substrate/contact induce non--zero
electron/hole charge density in the nanotube. This extra charge
density polarizes the substrate, which breaks the axial symmetry
of the nanotube. This effect is much larger than an electronic
structure perturbation caused by the lattice distortion which
may happen due to a van der Waals attraction to the
substrate\cite{avouris-vdw}. We will demonstrate that a direct
action of the \emph{uniform} external electric field is of minor
importance as compared to the \emph{nonuniform} field of surface
charges on the substrate. We will discuss a modification of our
theory of the subband splitting for a case of purely insulating
substrate in the last section.

\section{Perturbation Theory for Bandstructure Modification}
\label{sec:main}

\subsection{Splitting of SWNT subband due to transverse depolarization}
\label{sec:model}

To calculate the splitting and shift of the electron energy levels one needs to
know matrix elements of the perturbation potential between corresponding
wave--functions. In our case, the perturbation is a selfconsistent Coulomb
potential (operator in Heisenberg representation) which describes the
interaction between the probe electron and the extra charge density on the SWNT
and the polarization charge density on the substrate surface:
\begin{eqnarray}
 {\hat V}=e \int\limits_{-L/2}^{L/2}dZ
\int\limits_{0}^{2\pi} R d\beta
\left( \frac{\hat\sigma}{\sqrt{(z-Z)^2+(R
\cos\alpha- R \cos\beta)^2+(R \sin\alpha-R \sin\beta)^2}}
\nonumber \right.
\\
\label{coulomb}
\\ \nonumber
    \left.
+ \frac{\hat\sigma^*}{\sqrt{(z-Z)^2+(R \cos\alpha-
R \cos\beta)^2+(R \sin\alpha-2h-R \sin\beta)^2}}\right).
\end{eqnarray}
Both the probe electron and the nanotube surface charge are taken on a cylinder
of a radius $R$. Then $z$ and $\alpha$ are the electron coordinates in the
cylindrical coordinate system. $\sigma$ is the surface charge density. It does
not depend on the coordinate $Z$ along the nanotube because we assume the
translational invariance of the problem for clarity of derivation. Although, the
theory can be easily extended for the case of slow variation of $\sigma$ along
the axis. We will show later that one can drop dependence of $\sigma$ on the
angle $\beta$ along the circumferential direction in approximation of a
\emph{linear response} (in higher orders of perturbation theory a direct
transverse polarization must be taken into account\cite{yanli}). $\sigma^*$ is
an image charge density which is equal to $-\sigma$ for the metallic substrate.

The first term of Eq.(\ref{coulomb}) is the interaction with the charge density
on the nanotube, which coincides with the Hartree term for the
SWNT in vacuum (without charge injection). The second term in
Eq.(\ref{coulomb}) has also a simple physical meaning: this is
the energy of interaction of the electron with the image charge.
The separation between the SWNT axis and the surface of the
conductor is $h$. In case of the metallic substrate it is about the nanotube
radius, $R$, plus the van der Waals distance for graphite: $h\sim R+0.34$ nm.

The matrix element of the Coulomb operator (\ref{coulomb}) is calculated with
the wave--functions of a tight--binding (TB) Hamiltonian. We use envelope
wave--functions, obtained similarly to Ref.\cite{gonzalez}. This approach has
been widely used in the literature, so we skip details and give the
wave--functions in the one--band scheme ($\pi$ electrons only) in the form:
\begin{equation}
|\psi_{m,k,\zeta}\rangle=\frac{1}{\sqrt{2}}(|A\rangle +\zeta
c_{mk}|B\rangle)
e^{ikz} e^{im\alpha},
\label{wavefun}
\end{equation}
here index $m$ labels subbands of the SWNT electronic structure, $k$ is a
longitudinal momentum, these two are good quantum numbers (discrete and
continuum, respectively) for an ideal, long enough nanotube; $\zeta=\pm 1$ is a
pseudospin. (A pseudospinor vector is formed by a two--component wave--function
amplitude defined for two atoms in a graphite unit cell, A and B). Coordinate
along the tube is $z$, and $\alpha$ is the angle along the nanotube
circumference.

We assume that our potential is smooth at the scale of the single unit cell
(0.25 nm). Then, one may neglect transitions with the pseudospin flip
(transitions between sublattices). With use of the orthogonality relation
between the spinor components, it yields:
\begin{equation}
\langle m| V |n\rangle =
-\frac{8\pi e R\sigma}{|m-n|} i^{m-n}\left(\frac{R}{2h}\right)^{|m-n|},
\qquad m\neq n
\label{me1}
\end{equation}
\begin{equation}
\langle m| V |m\rangle =
 4\pi e R \sigma\log\left(\frac{2h}{R}\right),
\qquad m= n
\label{me2}
\end{equation}
\label{me}
where $\sigma$, the surface charge density, has to be defined
later in a selfconsistent way.

The Eqs.(\ref{me1},\ref{me2}) are obtained by a direct Fourier
transformation of (\ref{coulomb}) and
describe the energy level shift
when $m=n$ and the mixing of different subbands at $m\neq n$.
The most interesting term with $n=-m$ is the mixing between the
degenerate electron states within the same subband. By solving a
secular equation for the intrasubband mixing of the electron
doublet we obtain the splitting of the van Hove singularity at
the subband edge (Fig.\ref{fig:split}). The new subband energy
separation reads as:
\begin{equation}
\delta E_m=\frac{8\pi e R \sigma}{m}
\left(\frac{R}{2h}\right)^{2m}.
\label{split}
\end{equation}
Let us now calculate the injected/induced charge density $\sigma$ which will
allow us a numerical estimation for the $\delta E_m$ splitting.

\subsection{Charge injection due to the Fermi level shift}
\label{sec:fermi}

The Eqs.(\ref{me1}--\ref{split}) are written for the given charge density
$\sigma$ which has to be derived in this section. When the SWNT is placed in a
real device, one must consider the work function difference between the nanotube
and the contact or the conducting substrate and/or the external potential which
may be applied to the SWNT. The potential shifts the Fermi level in the
SWNT\cite{jetpl}. As a result of this the positive/negative charge is injected
into the nanotube:
\begin{equation}
\sigma =  \frac{e}{2\pi R}
\int\limits_{0}^{\mu(\sigma)} \nu (E) dE,
\label{sigma}
\end{equation}
here $\nu(E)$ is a bare one--dimensional DOS (independent of
$\sigma$ in a linear response theory); $\mu=\Delta W -e
\varphi^{xt} - e \varphi^{ind}(\sigma)$ is the shift of the
electrochemical potential of the SWNT (with respect to a charge
neutrality level $E=0$) which depends on the work function
difference, $\Delta W$, on the external potential,
$\varphi^{xt}$, applied between the nanotube and the reservoir
and on the potential $\varphi^{ind}$ induced by the charge
density of the nanotube, $\sigma$. This last term is
proportional to the intrasubband term ($m=n$) of the Coulomb
interaction given by the Eq.(\ref{me2}).

\begin{figure}[h]
\begin{center} \includegraphics[width=6cm]{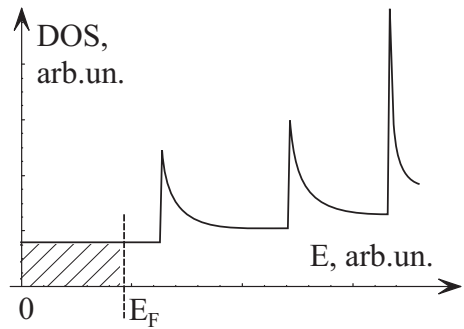} \end{center}
\caption{Schematic density of states (DOS) of a metallic SWNT.
First (massless) subband contributes to a constant DOS at the
$E=0$. When the Fermi level, $E_F$, is lower than the second
(massive) subband edge (which corresponds to the first peak of
DOS), an injected/induced charge is proportional to the shaded
area and is a linear function of $E_F$.}
\label{fig:dos}
\end{figure}

This selfconsistent equation for $\sigma$,
(\ref{sigma}), is readily solved
analytically if the electrochemical potential is below the
second subband edge. We follow
Ref.\cite{jetpl} in derivation of $\sigma$: the induced
potential is obtained by direct integration of the charge
density along the SWNT as in \cite{jetpl,ecsaluru}.  As it is
shown in Figure \ref{fig:dos}, for a metallic SWNT,
the charge is a product of the constant DOS, $C_Q$, and the
electrochemical potential, $\mu$.
Then the solution of
Eq.(\ref{sigma}) is as follows:
\begin{equation}
\sigma_A=\frac{\Delta W-e\varphi^{xt}}{2\pi R \, e \left(2 \log
(2h/R)+C^{-1}_Q \right) },
\label{intofdos1}
\end{equation}
and for a semiconductor SWNT, which has the DOS $\propto C_Q E
\theta(E-\Delta)/\sqrt{E^2-\Delta^2}$, the charge is as follows:
\begin{equation}
 \sigma_Z= \sigma_A
\frac{\sqrt{
\left( \frac{\Delta}{\Delta W-e\varphi^{xt}}\right)^2
\left( 4 \log^2(2h/R)-C^{-2}_Q \right)
+C^{-2}_Q}
- 2 \log (2h/R)}
{ 2 \log(2h/R)-C^{-1}_Q }
\theta(\Delta W-e\varphi^{xt}-\Delta).
\label{intofdos2}
\end{equation}
Here $\theta(x)$ is the Heaviside
unit step function and $\Delta$ is $1/2$ of the energy gap. We introduced a
quantum capacitance of the SWNT following Ref.\cite{jetpl}:
\begin{equation}
 C_Q = \frac{8 e^2}{3 \pi b \gamma}
\label{qmc}
\end{equation}
which is the one--dimensional analog of the quantum capacitance
proposed for a two--dimensional electron gas system by
Luryi\cite{luryi}. Here $b\simeq 1.4$\AA~ is the interatomic
distance, $\gamma\simeq 2.7$~eV is the hopping integral for the
graphite--like systems. We notice that despite the $\sigma_Z$,
as given by Eq.(\ref{intofdos2}), comes from a massive subband
(in contrast to $\sigma_A$, as in Eq.(\ref{intofdos1}) where the
lowest subband is massless, see in Fig.\ref{fig:dos}), the
linear dependence of $\sigma_{A/Z}$ on $\varphi^{xt}$ preserves
as long as the potential $\varphi^{xt}$ is large enough. This
reflects the fact that a classical one--dimensional charge
density is a linear function of a classical electrostatic
potential \cite{nano03}.

\section{Results and Discussion}
\label{sec:results}

\begin{figure}[h]
\begin{center} \includegraphics[width=6cm]{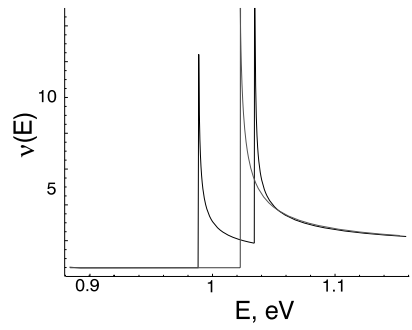} \end{center}
\caption{Density of states (DOS) of [10,10] armchair nanotube in
vicinity of first van Hove singularity (black color). Charge
injection in the NT due to work function difference (see the
text) results in a splitting of a doublet, which is clearly seen
as compared to bare DOS of neutral NT (light gray color).}
\label{fig:split}
\end{figure}

In the last section we obtained the selfconsistent expression
for the surface charge density as a function of the external
potential and the work function difference which may be
considered as a built--in potential. Substituting
Eq.(\ref{intofdos1}) into Eq.(\ref{split}) we obtain the
splitting of the degenerate subbands $|\pm m\rangle$ of the
metallic SWNT (when the Fermi level is within the first subband)
as follows:
\begin{equation}
\delta E_m=\frac{4
\left(\Delta W-e\varphi^{xt}\right)}{m
\left(2 \log (2h/R)+C^{-1}_Q \right)
} \left(\frac{R}{2h}\right)^{2m}.
\label{metsplit}
\end{equation}
The splitting decreases with $m$ exponentially, hence, the
effect is likely observable for the lowest degenerate subband.
Then, for the following parameters: the SWNT radius
$R\simeq 6.7$~\AA, the distance to the metal substrate
$h=10.1$~\AA, and the quantum capacitance $C^{-1}_Q \simeq
0.69$\cite{jetpl}, we obtain a numerical estimate for the
subband splitting $\delta E_1 \simeq 0.15 \left( \Delta
W-e\varphi^{xt}\right)$. Experimental data for the work function
of SWNTs scatters from 4.9 to 5.05 eV \cite{suzuki,shirashi}.
For the SWNT on the gold substrate we use as an estimate $\Delta
W\sim 0.3$~eV. In absence of the external potential, this work
function difference results in $\sim 46$~meV gap between two
split peaks of the density of states (Fig.\ref{fig:split}),
which is larger that $kT$ at room temperature. We also
calculated the contribution of all other subbands, which is
negligible in the splitting, but it shifts the doublet as a
whole. As a result, two new peaks in the Fig.\ref{fig:split}
appear not symmetrical with respect to the original DOS
singularity.

The splitting of $\pm m$ doublet is an analogue of a degenerate
level Stark effect for the nanotube in a multipole potential of
the image charge. The lower subband has $x$ symmetry and the
upper subband has $y$ symmetry (with corresponding
wave--functions $|x\rangle = 1/\sqrt{2}(|+m\rangle+|-m \rangle$
and $|y\rangle = 1/\sqrt{2}(|+m\rangle-|-m \rangle$) because of
an attraction energy of the electron to its image charge is
lower for the second combination.

We predict a similar effect for the semiconductor nanotube,
although, the total external potential causing the charge
density injection must be larger than one half of the gap in
this case. As we study in this paper only the effect which is
linear in the external potential, all high order terms in
Eq.(\ref{intofdos2}) have to be discarded.

\subsection{Dipole polarization correction}
\label{sec:dpc}

The charge injection in the nanotube may be readily achieved by applying
external electric field. One may naively argue that the external field itself
can break the bandstructure symmetry and result in some level splitting.
Although, it is correct statement in general, the direct splitting of the SWNT
orbital doublet $\pm m$ by the uniform electric field is forbidden by symmetry.
These degenerate states do not mix together due to the selection rules of the
problem. The matrix element for an intrasubband splitting in the uniform
external field ${\cal E}_{xt}$ equals zero by parity: $\langle m|e
{\cal E}_{xt} y|-m\rangle =0$.

In order to calculate the subband splitting in this case we have to compute the
charge injection, which is proportional to the applied field. The potential is
equal to $\varphi^{xt} ={\cal E}_{xt} h$, where $h$ is the distance between the
axis of the tube and the metal surface, which has to be substituted into
Eq.(\ref{metsplit}).

In Sec.\ref{sec:model} we assumed that the charge density $\sigma$ has no
dependence on the angular coordinate $\beta$ along the nanotube circumference.
This is an accurate approximation since a dipole (and higher multipole)
component of the $\sigma$ is small as compared to what is given by
Eqs.(\ref{intofdos1},\ref{intofdos2}). Let us prove this assumption for the
dipole polarization of the SWNT.

The non--uniform external potential causes a deviation of the surface density
from the uniform equilibrium value, $\sigma$, which is given by following
expression:
\begin{equation}
 \delta\sigma(\beta)=e\sum \limits_{i\neq j}
\frac{(f_i-f_j)\langle i|V|j\rangle}{E_i-E_j}
\langle j|\beta\rangle \langle \beta|i\rangle,
\label{angular}
\end{equation}
where $f_i$ are the occupation numbers, the matrix element $\langle
i|V|j\rangle$ is given by Eqs.(\ref{me1},\ref{me2}), $E_i$ are the energies of
subbands and $\langle \beta|i\rangle$ are corresponding wave--functions.

We define the nonuniform dipole part of the charge density of a SWNT as $
\delta\sigma_1\equiv \int \limits_{0}^{2\pi} \sin \beta \sigma(\beta) d\beta$.
Then, the dipole component of the surface charge is as follows:
\begin{equation}
 \delta\sigma_1= \frac{i e}{8\pi R^2} \sum
\limits_{i} \frac{(f_i-f_{i\pm 1})\langle i|V|i\pm 1\rangle}
{E_i-E_{i\pm 1}}.
\label{dipole}
\end{equation}

Let us remind that according to Eq.(\ref{me1}) $\langle i|V|i\pm
1\rangle = -i 8\pi R^2 e\sigma /(2h)$.

In case of the electrochemical potential equals zero (no charge
in the nanotube), the transverse polarization includes
transitions from the valence to conduction band only $\langle
v|V|c\rangle$ (the details of the calculation are presented
elsewhere\cite{yanli}). Here, we study an extra component of the
polarization, which is due to the induced charge density. Thus,
we need to consider only transitions from the levels above the
charge neutrality level, $E=0$, and below the Fermi level,
$E=E_F$, (which is the shaded area in Fig.\ref{fig:dos}). Hence,
the dipole polarization is proportional to the net charge
density $\sigma$, and the dipole charge density of the armchair
SWNT is given by the following expression:
\begin{equation}
 \delta\sigma_1=
 \frac{\sqrt{3} C_Q^2}{32 \pi}
\frac{\left( 2\pi
R\sigma_A\right)^2}{e}
\frac{R}{h}
\log \frac{2h}{R} \;
\propto {\cal E}^2_{xt}
\label{sigma1}
\end{equation}
where $C_Q\simeq 3.2$ is the dimensionless quantum capacitance. We single out
term $2\pi R\sigma_A$, which is the specific one--dimensional charge density of
the SWNT and proportional to the external potential and, thus, to the external
field.

The Eq.(\ref{sigma1}) shows that the effect of the transverse polarization on
the bandstructure is quadratic in the external field, in good agreement with a
plain dielectric response theory\cite{yanli}, while the effect of the image
charge is linear in ${\cal E}_{xt}$. Thus, the degenerate level splitting due to
the dipole component of the polarization will be less important than the
splitting due to a uniform component: $ \sigma_0\equiv \int_{0}^{2\pi}
\sigma(\beta) d\beta$, at least, in a weak field regime discussed in the
article. This proves post factum our assumption of $\sigma$ to be independent of
$\beta$.

\subsection{Depolarization at the insulator substrate}
\label{sec:diel}

For the sake of completeness we present here also a modification of our theory
to the case of a dielectric substrate. In this case the screening of the charge
density in the nanotube is weaker. It results from (i) underscreening of the
Coulomb interaction between the nanotube carriers and (ii) lower charge density
induced in the substrate. The second factor can be taken into account by
substituting an effective image charge density $ \sigma^*=\sigma
\frac{1-\varepsilon}{\varepsilon+1}$ in the second term of Eq.(\ref{coulomb}),
where $\varepsilon$ is the dielectric function of the substrate (in case of
highly conductive substrate it equals $-\infty$), instead of the bare image
charge density $-\sigma$. This results in substituting $\sigma^*$ in
Eqs.(\ref{me1}--\ref{intofdos2}) where appropriate.

Now, the fields of the image charge and the charge in the SWNT do not cancel
each other in contrast to the case of the metallic substrate. As a result, the
underscreening of the Coulomb interaction happens. This modifies the equations
for the energy level shift ( \emph{intrasubband} matrix elements as in
Eq.(\ref{me2})) and, thus, the electrochemical potential shift.
One must substitute $\log (2h/R)$ term everywhere by $\log
(2h/R) + 2/(\varepsilon-1)\log (L/R)$ where $L$ is the length of
the nanotube (or distance between metal leads to it). This
expression diverges with the length of the nanotube which
reflects the one--dimensional character of the Coulomb
interaction. These changes have to be made through
Eqs.(\ref{intofdos1}--\ref{metsplit}).

The first term of Eq.(\ref{coulomb}) does not appear in the
calculation of the \emph{intersubband} matrix elements as in
Eqs.(\ref{me1},\ref{split}). Hence, no additional correction is
required in the equations (\ref{angular}--\ref{dipole}) of the
last section.

We assumed in this paper that the perturbation theory in a linear approximation
in $\mu$ (or equivalently in $\sigma$) is applicable. Restrictions which may
follow from this assumption are as follows. The external potential has to be
small. We neglect here the dipole term in the induced charge density (and higher
multipoles as well). It is equivalent to a weak intersubband mixing which
assumption may not hold for wide nanotubes or strong external fields.  The
effect of the strong field on the bandstructure is discussed
elsewhere\cite{yanli}. In this paper we used the Eq.(\ref{sigma}) for the
equilibrium charge density in the SWNT. One may consider transport devices on
equal basis, as long as the charge of the nanotube is still given by the
quasi--equilibrium charge density. However, for non--zero current flowing
through the nanotube, one must use an expression for the charge density which
differs from Eq.(\ref{sigma}) (to be discussed elsewhere \cite{nano03}).

\section{Conclusions}
\label{sec:concl}

In summary, we have developed a microscopic quantum mechanical theory for a
charge transfer between a SWNT and a conductive substrate (and/or metallic
leads). This charge injection results from a natural work function difference
between the nanotube and the substrate or/and from an external potential applied
between those. A surface charge density of the SWNT is calculated
selfconsistently within an envelope function formalism of tight--binding
approximation.

We demonstrated for the first time that the influence of this charge transfer on
the electronic structure of the SWNT is not negligible for typical material
parameters of the problem. Because of the breaking of the axial symmetry of the
system, the SWNT DOS changes qualitatively: degenerate subbands $\pm m$, where
$m\neq 0, n$, split. It has a simple physical interpretation --- the electrons
with $x$ and $y$ polarizations are no longer equivalent as their attraction to
the substrate is different. This effect can be related to a degenerate level
Stark effect with an appropriate choice of external field of the image charge.
The gap between the new $x$ and $y$ subbands is constant in $k$--space (for the
external field which is uniform along the tube) so it shows up dramatically at
the subband edge. The van Hove singularity splits, and the distance between two
peaks of the DOS is about 46 meV for the [10,10] armchair SWNT on the gold
substrate.

We obtained analytical expressions for matrix elements of the image charge
field, which yields the mixing of different subbands and can be used to describe
the level anticrossing (to be discussed elsewhere). Same matrix elements enter
the expression for the multipole polarizabilites of the SWNT. We estimated a
major contribution to the dipole polarizability of the armchair SWNT, which
comes from intraband transitions for non--zero charge injection. The analytical
expression for the dipole component of the surface charge density is shown to be
proportional to the square of the external potential and, hence, appears in the
second order of the perturbation theory which corroborates post factum our
assumption of uniformity of the induced/injected charge along the SWNT equator.

We show that the modification of our theory to the case of semiconductor
substrate is straightforward. The analytical expressions for the van Hove
singularity splitting and induced charge density are obtained.

\textbf{Acknowledgments.}
S.V.R.~acknowledges support of DoE through grant
DE-FG02-01ER45932, NSF through grant ECS-0210495
and Beckman Fellowship from the Arnold and Mabel Beckman Foundation.
A.G.P.~is grateful to the Beckman Institute for hospitality
during his work in Urbana. Authors are indebted to Professor K.
Hess for valuable discussions.

\begin{figure}[h]
\begin{center}\includegraphics[width=6cm]{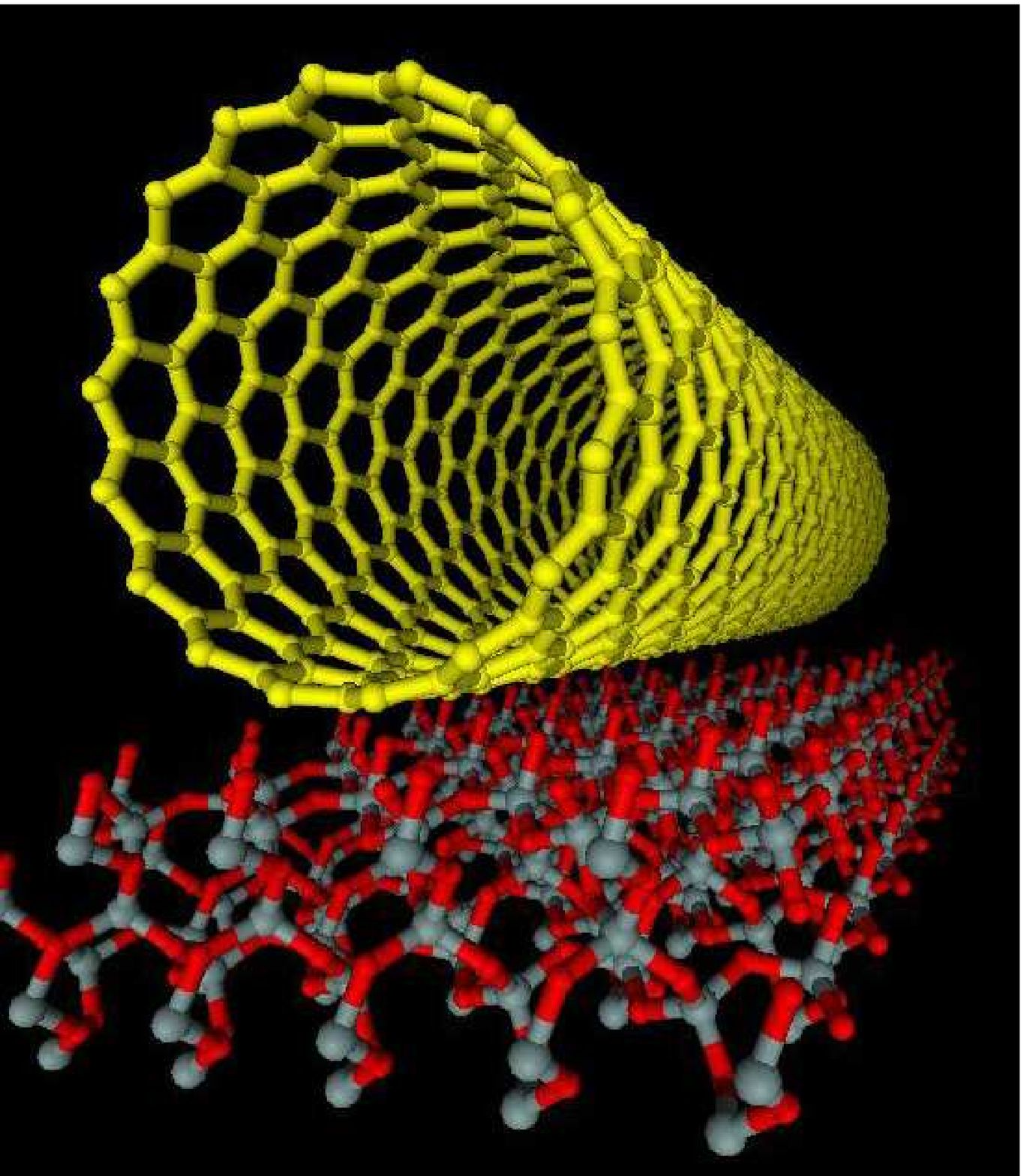}\end{center}
\caption{Image for table of contents: Zigzag [17,0] nanotube on a
surface of SiO$_2$ substrate. Geometry has been relaxed with
Molecular Mechanics.}
\label{fig:view}
\end{figure}


\begin{thebibliography}{99}
\bibitem{iijima} S. Iijima, Nature, \textbf{354}, 56, 1991.
\bibitem{appl} Ph.G. Collins, Ph. Avouris, Sci.Am. \textbf{12}, 62 , 2002.
\bibitem{louie} S.G.~Louie, \emph{in Carbon Nanotubes: Synthesis, Structure,
Properties, and Applications}, Eds.: Dresselhaus, M. S., Dresselhaus, G., and
Avouris, P.; Springer-Verlag, Berlin. 113--146, 2001.
\bibitem{jetpl}  K.A. Bulashevich, S.V. Rotkin, JETP Letters \textbf{75}(4),
205, 2002.
\bibitem{ecsaluru} S.V. Rotkin, V. Shrivastava, K.A. Bulashevich, N.R. Aluru,
Int. Journal of Nanoscience \textbf{1} (3/4), 337,
2002.
S.V. Rotkin, K.A. Bulashevich, N.R. Aluru, in Procs.--ECS \textbf{PV 2002--12},
P.V. Kamat, D.M. Guldi, and K.M. Kadish, Eds.: ECS Inc., Pennington, NJ, USA.
512, 2002.
\bibitem{tersoff} F. Leonard, J. Tersoff, Appl. Phys. Letters \textbf{81}
(25), 4835, 2002.
\bibitem{damnjan} T. Vukovic, I. Milosevic, M. Damnjanovic, Phys.Rev.
\textbf{B 65}(4), 5418, 2002.
\bibitem{avouris-vdw} T. Hertel, R. E. Walkup, and P. Avouris, Phys.Rev.
\textbf{\bf B 58}, 13870, 1998.
\bibitem{yanli} Y. Li, S.V. Rotkin, U. Ravaioli, Nano Letters \textbf{3} (2),
183, 2003.
\bibitem{gonzalez} J. Gonzalez, F. Guinea, MAH. Vozmediano, Nuclear Physics B.
\textbf{406} (3), 771, 1993. D.P. DiVincenzo, E.J. Mele, Phys.Rev.
\textbf{B 29}(4), 1685, 1984.
\bibitem{luryi} S. Luryi, Appl. Phys. Letters \textbf{52} (6), 501, 1988.
\bibitem{nano03} S.V.~Rotkin, H.~Ruda and A.~Shik, submitted, 2003.
\bibitem{suzuki} S. Suzukia, Ch. Bower, Y. Watanabe, O. Zhou, Appl. Phys. Lett.
\textbf{76} (26), 4007, 2000.
\bibitem{shirashi} M. Shiraishi, M. Ata, Carbon \textbf{39}, 1913, 2001.
\end{thebibliography}
\end{document}